\documentclass[preprint]{elsarticle}
\usepackage[numbers]{natbib}
\usepackage[tight,TABTOPCAP]{subfigure}

\usepackage{graphicx}
\usepackage{xcolor}
\usepackage{mdframed}

\newcommand{\ud}{\mathop{\mathrm{{}d}}\mathopen{}}

\newcommand{\be}{\begin{equation}}
\newcommand{\ee}{\end{equation}}
\newcommand{\ba}{\begin{eqnarray}}
\newcommand{\ea}{\end{eqnarray}}
\newcommand{\ban}{\begin{eqnarray*}}
\newcommand{\ean}{\end{eqnarray*}}

\newcommand{\demi}{\frac{1}{2}}

\newcommand{\re}{\mbox{Re}}
\newcommand{\norm}[1]{\left|#1\right|}

\begin{document}
\markboth{Ting}{Dipole light-harvesting antennas}


\title{Proposal for verifying dipole properties of light-harvesting antennas}








\author[df]{Julian Juhi-Lian Ting}
\ead{juhilian@gmail.com}
\ead[url]{http://amazon.com/author/julianting}
\address[df]{De-Font Research Institute, Taichung 40344, Taiwan, R.O.C.}
\date{\today}

\begin{abstract}
For light harvesters with a reaction center complex (LH1-RC complex) of three types, 
we propose an experiment to verify our analysis based upon antenna theories 
that automatically include the required structural information.
Our analysis conforms to the current understanding of light-harvesting antennas in that we can explain known properties of these complexes.
We provide an explanation for 
the functional roles of the notch at the light harvester,
a functional role of the polypeptide called PufX or W at the opening,
a functional role of the special pair, 
a reason that the cross section of the light harvester must not be circular,
a reason that the light harvester must not be spherical,
reasons for the use of dielectric bacteriochlorophylls instead of conductors to make the light harvester,
a mechanism to prevent damage from excess sunlight,
an advantage of the dimeric form, and
reasons for the modular design of nature.
Based upon our analysis we provide a mechanism for dimerization.
We predict the dimeric form of light-harvesting complexes is favoured under intense sunlight.
We further comment upon the classification of the dimeric or S-shape complexes.
The S-shape complexes should not be considered as the third type of light harvester but simply as a composite form.


\end{abstract}

\begin{keyword}
dipole antenna \sep S-shape antenna \sep light harvester \sep  electrodynamics \sep photosynthesis
\PACS{wave optics 42.25.-p ;  biomolecules 87.15.-v}
\end{keyword}

\maketitle


\section{Introduction}

Photosynthesis is divided into light-dependent reactions and the Calvin cycle (carbon fixation reactions) \citep{Nelson2013,Berg2015}.
The primary step of the light-dependent reaction includes energy transfer and electron transfer.
The Calvin cycle is a sequence of  chemical processes, whereas the energy transfer step might be analyzed physically.

To consider the interaction of light and a single atom we can approximate the system as a two-level system.
However, if the number of atoms grows and the system becomes extended in space, the interference from individual effects will include time-lag.
Biological systems often fall into this later case.
A theory to sum-up individual effects of light-matter interaction is called antennas theory.
In an antenna theory the most important information is the geometry of the antenna which will produce the time-lag.
A proposal has appeared in which DNA is considered as a fractal antenna when placed in electromagnetic fields\citep{Blank2011}.


Beginning about 1995, a reasonably complete picture of the bacterial light-harvesting (LH) systems has been acquired  ~\citep{Kuhlbrandt1995a,Hu1997,Cogdell2006}.
Many such structures have been subsequently analyzed \citep{Cogdell1999}.
The light harvesters with a reaction center complex (LH1-RC) is particularly interesting 
as it is a fully fledged antenna, even without LH2. 
Therefore, the LH1-RC complex serves as a minimum model we should consider.

\begin{figure}[tbp]
\begin{tabular}{llcc}
\hline 
protein & PDB ID & symmetry & cartoon \\
\hline \\

{\it Tch. tepidum}  \citep{Niwa2014}              & 4V8K & C16 &
\begin{minipage}[c]{0.15\textwidth}\includegraphics[width=\textwidth, angle=0]{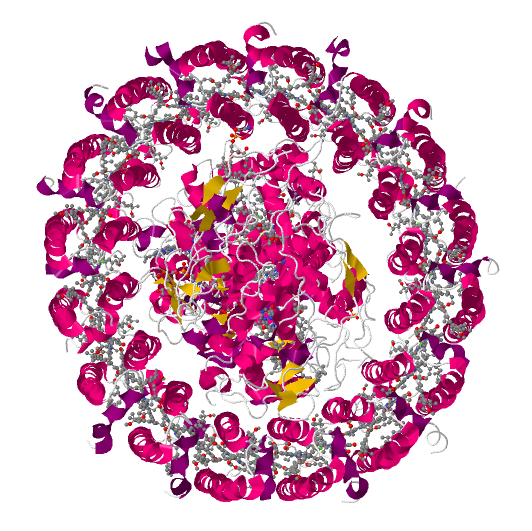}\end{minipage}  \\

{\it Rps. palustris}  \citep{Roszak2003}              & 1PYH &  &
\begin{minipage}[c]{0.15\textwidth}\includegraphics[width=\textwidth, angle=0]{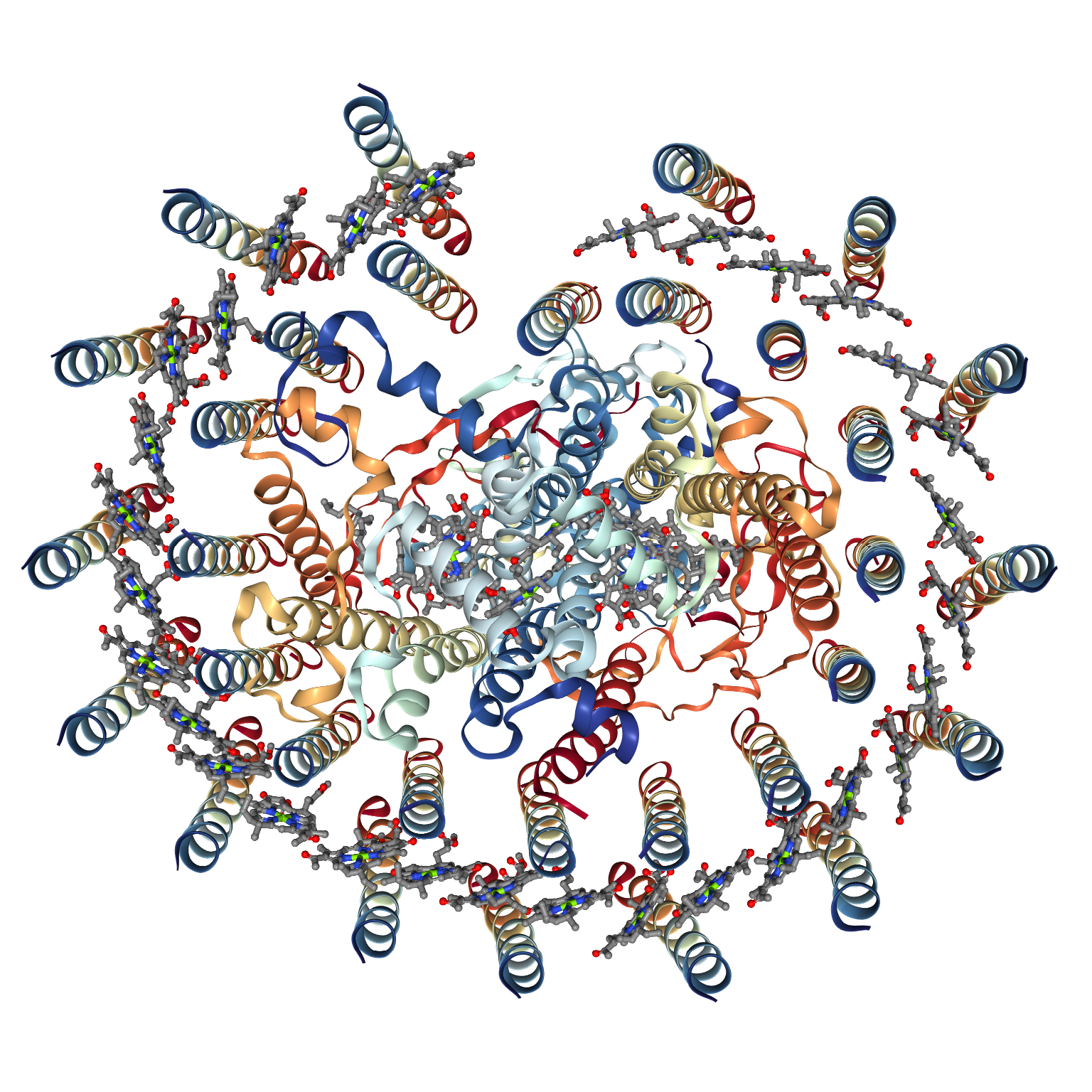}\end{minipage}  \\

{\it Rba. sphaeroides}  \citep{Qian2013}              & 4V9G &  &
\begin{minipage}[c]{0.15\textwidth}\includegraphics[width=\textwidth, angle=0]{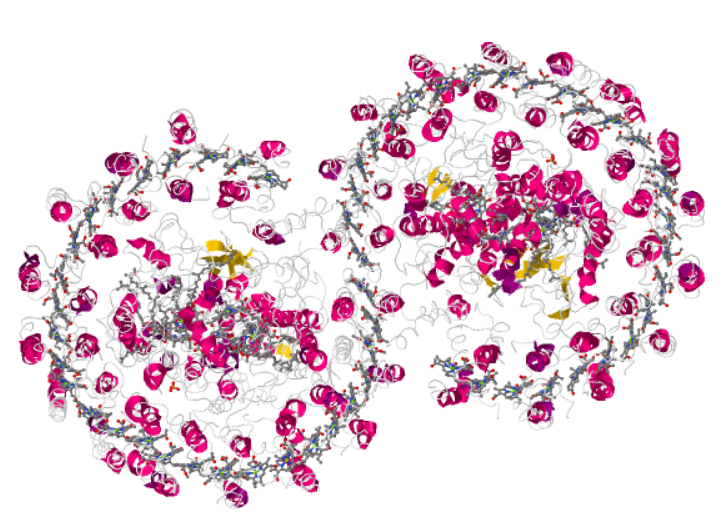}\end{minipage}  \\

\\
\hline 
\end{tabular}
\caption{Three types of LH1-RC complex.
The molecules are identified by the PDB ID assigned by Protein Data Bank http://www.rcsb.org/pdb/
}
\label{more}
\end{figure}

Such antenna of three basic types have been discovered:
\begin{itemize}
\item

The most common form of LH1-RC complex exists as monomeric form with RC surrounded by a closed elliptical ring
such as for
{\it Blastochloris viridis} \citep{Jay1984}, 
{\it Phaeospirillum (Phs.) molischianum}; previous name {\it Rhodospirillum (Rsp.)} \citep{Boonstra1994,Imhoff1998}, 
{\it R. rubrum} \citep{Fotiadis2004}, 
{\it Rhodobacter (Rba.) veldkampii} \citep{Busselez2007,Liu2011a},
{\it R. capsulatus} \citep{Crouch2012}, 
{\it R. vinaykumarii} \citep{Crouch2012}, 
and 
{\it Thermochromatium (Tch.) tepidum} \citep{Niwa2014}. 

\item

In {\it Rhodopseudomonas (Rps.) palustris}, the complex is found in a monometric form that contains an opening \citep{Roszak2003}.
The cross section of this LH1-RC complex is elliptical; 
the RC is surrounded with an incomplete double ring of helices. 
Some species, such as {\it R. sphaeroides}, have a polypeptide component termed "W" is found at the opening, 
similar to the PufX polypeptide in several {\it Rba.} species
for which a precise function has long been debated\cite{Hsin2010,Semchonok2012,Crouch2012,Qian2013,Olsen2017}.
The elliptical LH1 complex has an outer long axis of length 110 \AA ~and a short axis of length 95 \AA ~;
the greatest dimension of the inside of this LH1 is 78 \AA .
A gap of 4 \AA ~ is found between the RC and the ring, as the RC has a long dimension of 70 \AA ~ and 
the orientation of the long axis of the LH1 ellipse coincides with the long axis of the RC.
The height of the cylindrical (more precisely toroidal) shape of the molecule can be obtained with software such as Jmol or PyMOL, 
using the data from PDB, to be about half the width of the molecule.

\item

A dimeric form exists in {\it R. sphaeroides} \citep{Jungas1999,Scheuring2004,Qian2005,Sturgis2008,Crouch2012},
{\it R. blasticus}\citep{Scheuring2005}, {\it Rhodobacter changlensis} and {\it Rhodobacter azotoformans}\citep{Crouch2012}.

\end{itemize}
We show the structures of a number of these complexes in Figure \ref{more}.
In some species, such as {\it Rhodobaca bogoriensis}, monomeric and dimeric forms have been shown to coexist\citep{Semchonok2012}.

The standard model to consider energy transfer within light-harvesting complexes in the photosynthesis community is the exciton theory.
For instance, Komatsu {\it et al.} used exciton theory to calculate the absorption spectra and energy transfer rates
of an array of LH2 \citep{Komatsu2015}.
They approximated the ring as an array of dipoles.
The problem of this model is it involves a calibration procedure, which is {\it ad hoc}, even though the authors claim their calculation is
{\it ab initio}.

We address the three forms mentioned, in particular the dimer form from a classical electrodynamic point following the analysis of non-reciprocal properties in
a previous paper\cite{Ting2017}, in the following.
Our analysis takes the structural information into consideration and requires no {\it ad hoc} parameter.
We hence propose an experiment to verify our theory.

\section{Loop Antennas}
\label{loop}

\begin{figure}[tb]
\begin{center}

\mbox{
\subfigure[]{\includegraphics[width=0.25\textwidth, angle=0]{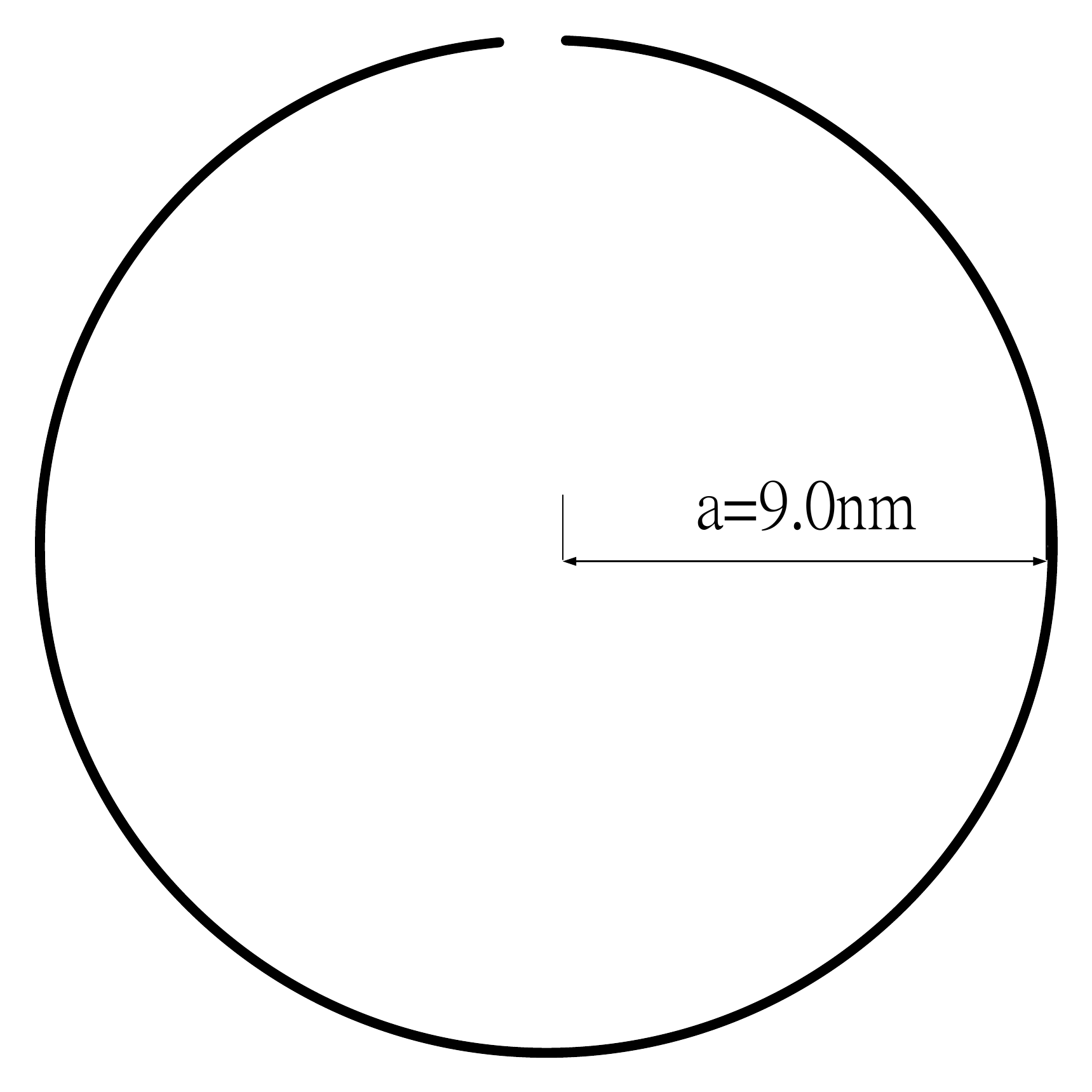}}
\subfigure[]{\includegraphics[width=0.25\textwidth, angle=0]{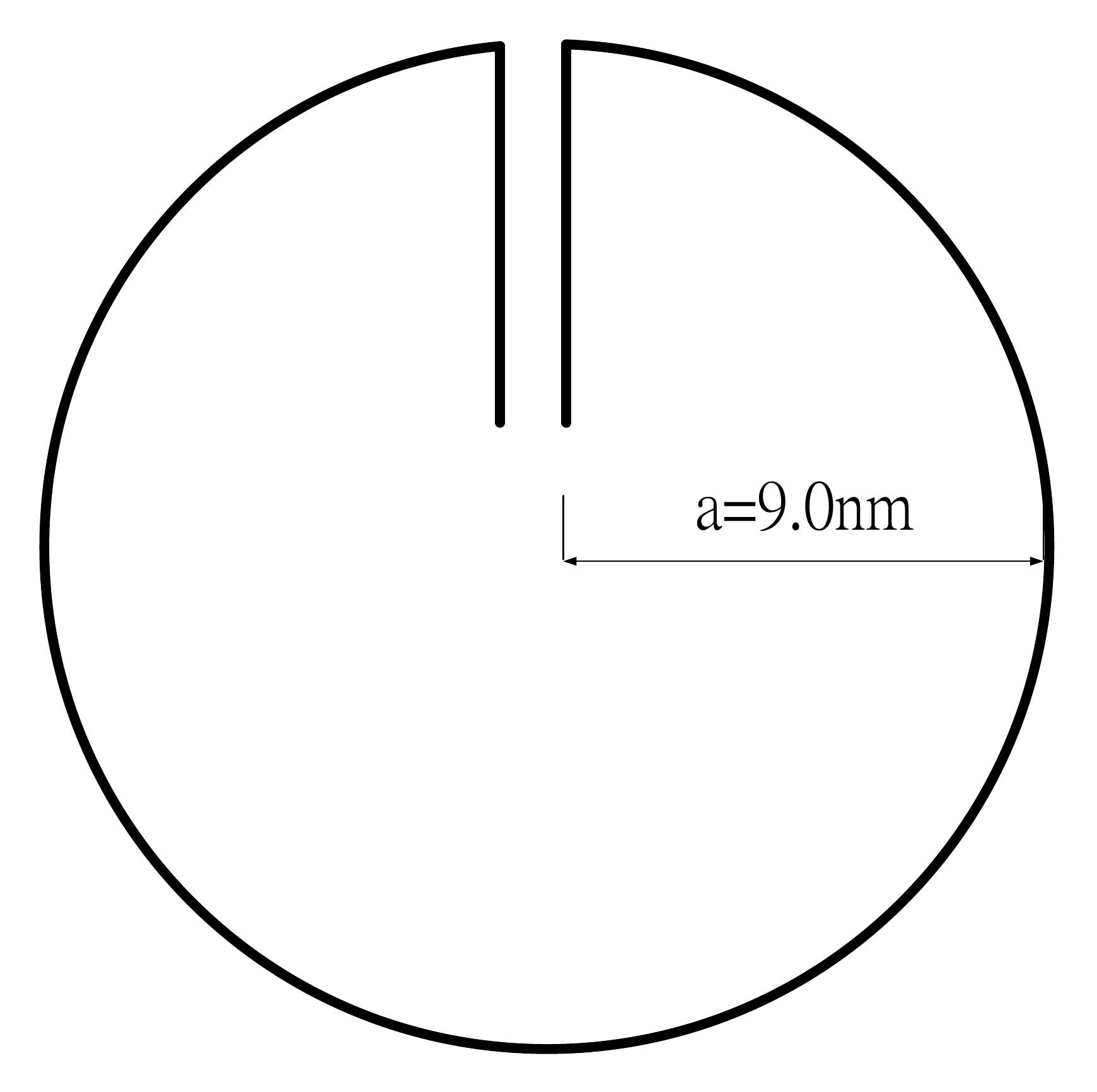}}
}
\caption{
Antennas of two idealized shapes.  
According to  experimental data, 
the exterior length of the long axis of the {\it Rps. palustris} LH1-RC complex is 110 \AA ,
while the short axis is 95 \AA; 
the longest dimension of the inner LH1 is 78 \AA .
~We choose $90$ \AA ~as the average value in our figures.
(a) is a simple loop antenna, while (b) is called a loop antenna with line feed.
The opening can be filled with other material to adjust the resonance frequency, whereas the LH1 has a PufX/W over there. 
The feed line, though resembling the special pair in the RC,
modifies only slightly the resonant frequency, but the notch is essential.
}
\label{models}
\end{center}
\end{figure}

Two simplified shapes shown in the figure resemble the {\it R. palustris} (1PYH) LH1-RC complex
shown in Figure \ref{more}. 
Figure \ref{models} (a) is simple and can be solved with algebraic methods, 
whereas figure \ref{models} (b), although resembling the LH1-RC complex more closely, 
has a resonant frequency only slightly modified from that of figure \ref{models} (a).
We hence consider only figure \ref{models} (a).
The notch is essential not only for concentrating but also to take out the energy received by the $LH1-RC$ complex\cite{White2009}.
The  molecule of 1PYH shown in Figure \ref{more} also has a notch.
Chemists describe the function of the opening as enabling the passage of quinones (charge carriers) to the RC \citep{Jungas1999}.
Without this opening, such as the molecule 4V8K, other mechanisms have to be employed to retrieve the energy received.
In engineering the opening can be filled with a spacer material to adjust the resonance frequency, which is interpreted
as impedance matching in antenna theory\cite{Monticone2017}, while in biology a polypeptide called PufX or W is often found over there.
Our analysis here show that PufX/W and the notch arise from distinct physics; one does not  imply the other.

In engineering, antennas of such shapes are well studied: 
they are called loop antennas and loop antennas with line feed, respectively \citep{Kraus1988}.
The line feed resembles the special pair from the light harvester to the reaction center.

Loop antennas can be further divided into two categories depending upon their size relative to the wavelength of operation.
If an antenna has a radius smaller than the wavelength of operation, 
it is called a small-loop antenna; 
otherwise, a resonant-loop antenna.
As the wavelength, $\lambda$, of operation of LH1 or LH2, $800-900~nm$, 
is much larger than the radius of the antenna, $9.0 ~ nm$, the light-harvesting antenna is a small-loop antenna, more precisely deep sub-wavelength antenna,
which has a small radiation resistance and a large reactance; 
its impedance is hence difficult to match with that of the transmitter.
As a result such antennas serve mainly as a receiving antenna for which an impedance mismatch loss can be tolerated.
A small-loop antenna is equivalent to a short-dipole antenna of which the receiving (radiation) pattern has a toroidal shape, 
with electric and magnetic fields interchanged, and thus serves as a magnetic dipole with the direction of dipole orthogonal to the loop plane.

There are a complete way and a simple way for arriving at the electromagnetic properties of such an antenna.
We begin with the complete way.

\begin{figure}[tb]
\centering
\mbox{
\subfigure[]{\includegraphics[width=0.5\textwidth, angle=0]{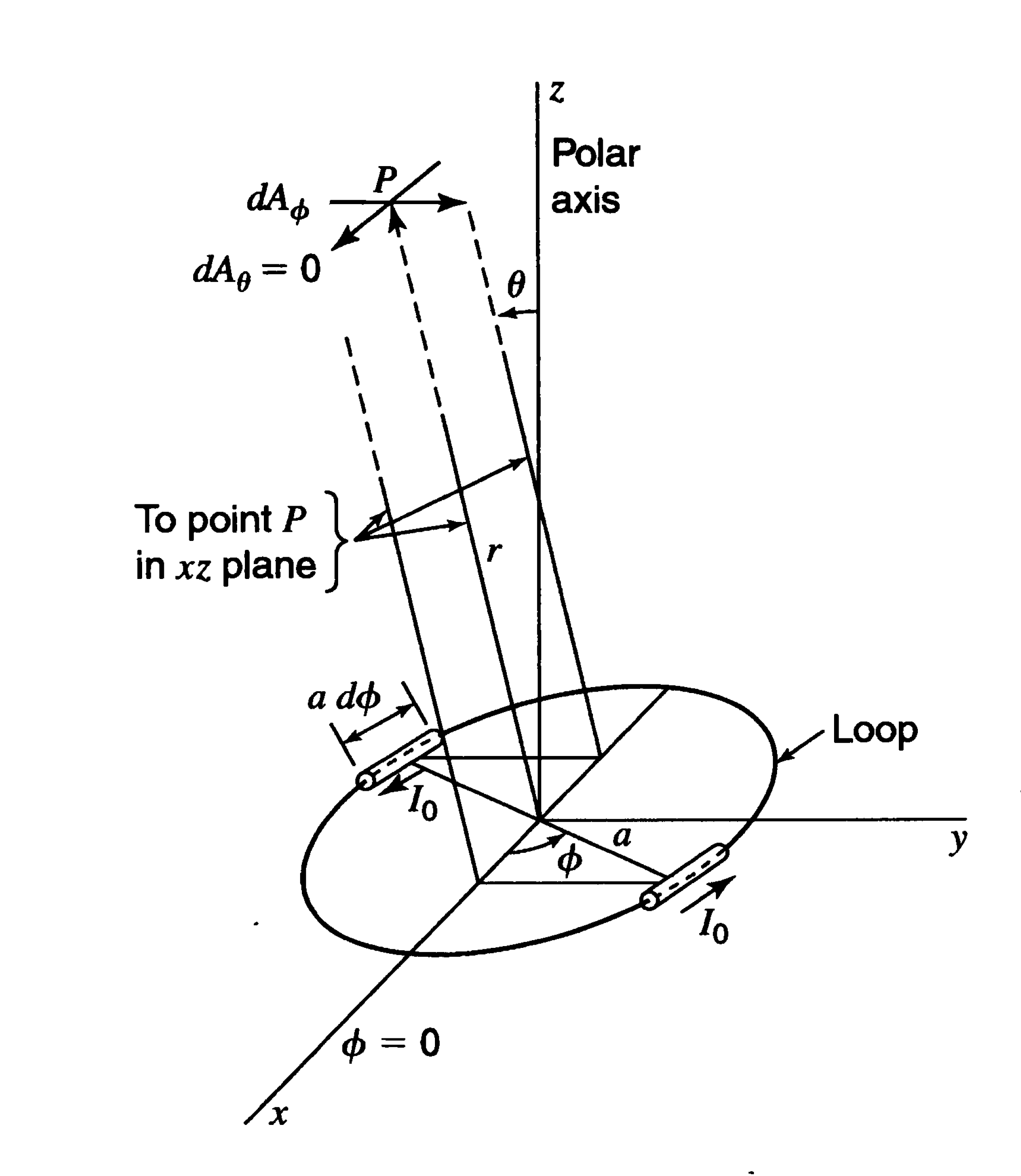}}
\subfigure[]{\includegraphics[width=0.5\textwidth, angle=0]{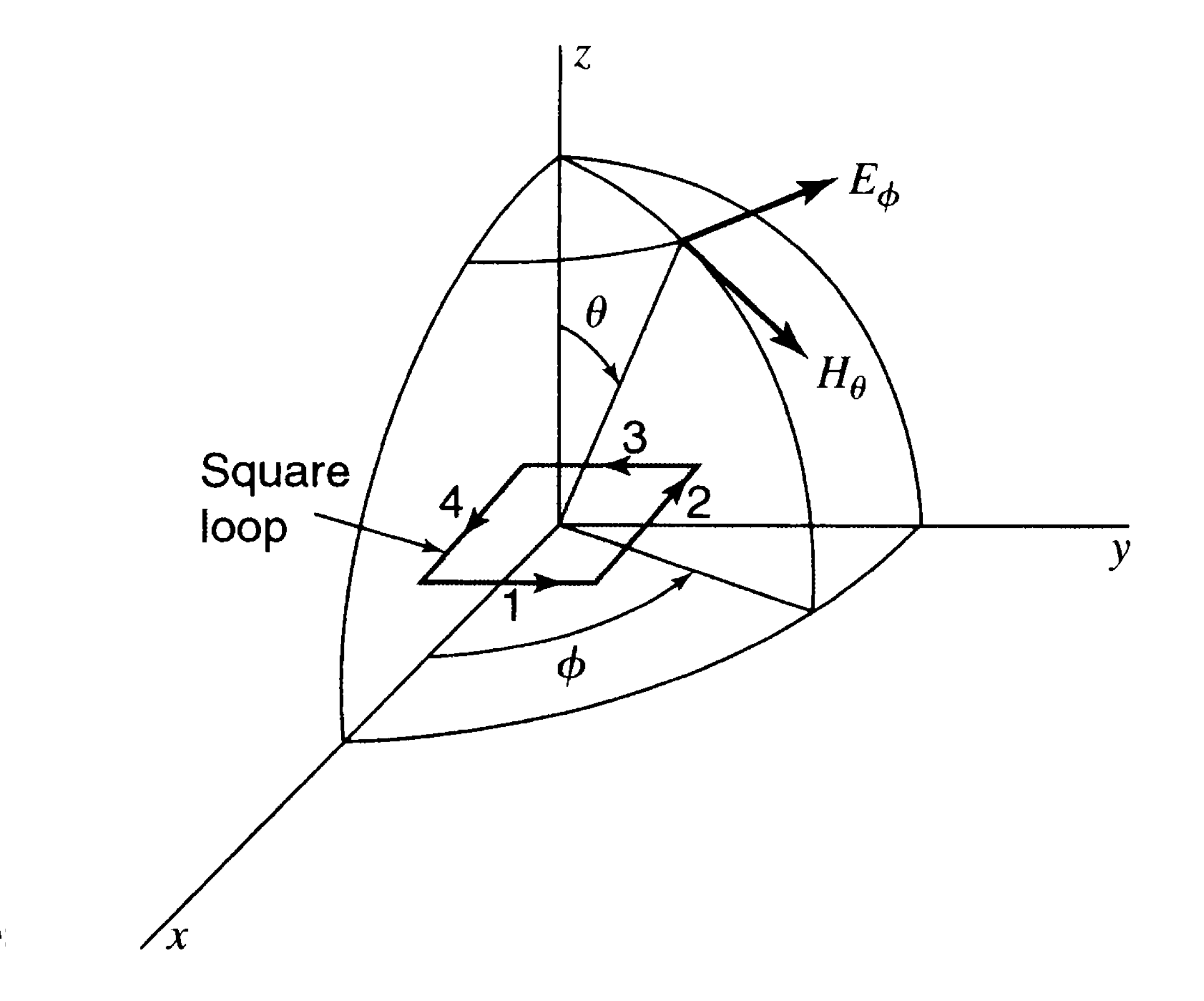}}
}
\caption{The coordinate system uses.}
\label{geometry}
\end{figure}

Let the radius of the loop located at the origin be $a$, the plane of the loop be $x-y$, and the angle from the $x-$axis be $\phi$. 
If the current $I$ around the loop is uniform and in phase,
the only component of the vector potential is $A_\phi$, as shown in Figure \ref{geometry} (a).
The infinitesimal value of $A_\phi$ at a point away from the loop by distance $r$ caused by two diametrically opposed infinitesimal dipoles is
\be
\ud A_\phi = \frac{\mu \ud M}{4 \pi r} \,,
\ee
in which $\ud M = 2 j [I] a \cos \phi [\sin (2 \pi a \cos \phi \sin \theta / \lambda) ] \ud \phi$,
$\theta$ is the angle relative to the vertical axis
through the center of the loop, 
and
$[I]= I_0 \exp{\{j \omega [t-(r/c)]\}}$ is the retarded current on the loop with $I_0$ being its maximum value.
After integration we obtain
\be
A_\phi = \frac{j \mu [I] a}{2r} J_1 (\frac{2 \pi a \sin \theta}{\lambda} ) \,,
\ee
in which $J_1$ is a Bessel function of first order.

As the source of sunlight is remote, we consider the far-field effects.
The far electric field of the loop has only a $\phi$-component $E_\phi = -j \omega A_\phi$ that is in the plane of the loop. Therefore,
\be
E_\phi = \frac{120 \pi^2 a [I]}{\lambda r} J_1 (\frac{2 \pi a \sin \theta}{ \lambda}) \,.
\label{complete1}
\ee
The corresponding magnetic field in free space is
\be
H_\theta = \frac{\pi a [I]}{\lambda r} J_1 (\frac{2 \pi a \sin \theta}{ \lambda}) \,.
\label{complete}
\ee

The second method is to decrease the infinite number of dipoles used in the preceding method into four short linear dipoles as follows.

Let the area of the antenna be $A$, which is commonly called the aperture of the antenna, and the length of the dipoles be $d$, as shown in Figure \ref{geometry} (b). 
Hence
\be
d^2 = \pi a^2 \equiv A
\ee
The far electric field is
\be
E_\phi = \frac{120 \pi^2 [I] \sin \theta }{r} \frac{A}{\lambda^2} \,.
\label{simple1}
\ee
The term $A / \lambda^2$ is a pure ratio: it is the aperture in terms of wavelength.
The magnetic field is obtained on dividing by the intrinsic impedance of the medium, i.e. 
\be
H_\theta = \frac{E_\phi}{120 \pi} = \frac{\pi [I] \sin \theta }{r} \frac{A}{\lambda^2}
\label{simple}
\ee
in a vacuum.

Eq. (\ref{simple1}) is a special case of Eq. (\ref{complete1}), 
just as Eq. (\ref{simple}) is a special case of Eq. (\ref{complete}), 
because, for small arguments of the first-order Bessel function, $J_1 (x) \approx x/2$\citep{Arfken2001}.

The (radiation or receiving) resistance at the loop terminals can be obtained from
\be
P = \frac{I_0^2}{2} R
\ee
in which $I_0$ is the maximum current on the loop and $R$ is the resistance.
The total power $P$ is obtainable on integrating the Poynting vector
\be
S = \demi \norm{H}^2 \re Z
\ee
over a large sphere, in which $Z$ is the impedance of the medium.
The resistance thus obtained for a small-loop antenna is proportional to $1/\lambda^4$.

This analysis is a direct consequence of Maxwell’s equations, which have no restrictions on the range of applicable frequencies.
The assumption behind these calculation is that the 
conducting wire making up the loop is infinitesimally thin
so that the current within the wire can be assumed to be uniform, which is irrelevant to our analysis.
A numerical calculation of finite-thickness loop nanoantenna is done by Locatelli\cite{Locatelli2011}.

\section{dipole antennas}

Biologists describe the third form of a LH1-RC complex as a dimer, which conveys no further meaning except it is formed from two parts.
Semchonok et al. interpreted Table 3 and Figure 5 of Crouch and Jones to signify that,
for species of which both monomers and dimers exist, 
the ratio between them depends on the light intensity, even though the latter authors conducted no related experiment \citep{Semchonok2012,Crouch2012}.

However, as we show in section \ref{loop}, a monomer can be described by a small-loop antenna, 
which has a radiation character as a dipole.
Although a cylindrical antenna is a monopole, dielectric resonator antennas operating at their fundamental modes always
radiate like a magnetic dipole independent of their shapes.
If two dipoles are put together, they become a quadrupole.
A quadrupole is less efficient in radiation or receiving than a dipole, 
whereas a dipole is less efficient than a monopole\cite{Jackson2007}. 
Accordingly, we predict that a dimeric form is preferable under intense sunlight.
Furthermore, under even greater intensity or at the same intensity but with a smaller ratio a tetrameric form might appear.

To verify the theory above we propose an experiment:
\begin{mdframed}[backgroundcolor=blue!3] 
\begin{itemize}
\item Grow the cells under light of varied controlled intensity.
\item Take images under the microscope.
\item Count the number of monomers and of dimers.
\item Plot the ratio of the number of the monomers to that of dimers and 
the ratio of the number of dimers to the number of tetramers against the light intensity 
to determine whether these ratios decrease when the light intensity increases.
\end{itemize}
\end{mdframed}

The experiment is as simple as Robert Hooke's experiments to observe a cell except that our domain is much smaller, on a nanometer scale. 
The numbers of dipoles and quadrupoles are measured simply on counting in the photographs taken.
An image-processing program can be written to undertake this task, 
which is a standard technique in image processing, 
if many images are to be treated, but a few photographs suffice for our purpose. 

\section{Discussion}

The analysis in section \ref{loop} is all that is required for the theoretical background behind the experiment proposed.
A few more words might be better:

The antenna must work with symmetry breaking of the electric field in space to function\citep{Sinha2015}, 
which is the third reason for the mysterious opening at the LH1-RC complex and explaining the non-circular shape of the cross section observed.
Poincar\'e-Brouwer theorem\citep{Brouwer1911} further restricts the shape of the antenna to be toroidal instead of spherical, 
which is the second lesson from nature about solar light harvesting mentioned by previous authors\citep{Scholes2011}.
Anapole radiation for non-reciprocity also favor toroidal shape\citep{Miroshnichenko2015,Ting2017}.
The non-spherical requirement is not only depicted at a smaller scale as for the light-harvesting complex discussed here but also
at a larger scale as in chromatophores\cite{Chandler2009}.

Another possibility of doubt about the validity of the formalism presented 
might concern whether the model above is applicable for a
light harvester that is composed of dielectric material, 
whereas the antenna in section \ref{loop} is made of conducting material.
Could a dielectric serve as an antenna?
Antennas are simply devices mediating between a source, or a
receiver, and the electromagnetic field.
At the frequency of visible sunlight the electrons inside metals have
difficulty following the rapid oscillation of the electromagnetic waves; a dielectric works more efficiently.
The antennas inside our mobile telephones are mostly dielectric ones.
Some dielectric nanoantennas have been fabricated
recently\cite{Devilez2010,Zou2013}.

Unlike a metallic antenna, a dielectric one requires no electron flow inside its body.
It works more or less like an acoustic resonator; the electrons are bouncing inside the cavity.

To use a dielectric as a receiver of sunlight has several other advantages:
\begin{itemize}
\item The size of the antenna is smaller for a dielectric antenna than for a metallic one.
\item Dielectric receivers immunize the antennas from damage of high power\citep{Hsu2007,Hsu2007a}, 
which is generally attributed to the carotenoids involved, but the FMO complex has no carotenoid\citep{Cogdell2000}.
\item The bandwidth can be increased on adjusting the size of the cylinder or the size of the hole at the centrer of the cylinder.
\end{itemize}

When sunlight (an electromagnetic wave) excites the resonance of an antenna, 
a mode-field pattern is built inside the structure.
The location of each module of bacteriochlorophyll corresponds to
a mode of the electric field of the antenna\citep{Yu2012,Soren2014}.
The classical F\"orster theory is a local resonance-energy transfer of a
kind that confirms this interpretation.
Furthermore, a ring composed of discrete subunits works better than a
continuous ring, 
as a ring of nanoparticles corresponds to an interconnection of inductors
interleaved by capacitors that guides the flow of displacement current
better\cite{Alu2008a}.
Such a split-ring resonator is used commonly in microwave engineering.
More are discussed in a subsequent paper\cite{Ting2018}.

\section{Summary}

In this work, we seek an approach to consider a light-harvesting antenna as a device to receive electromagnetic waves.
We tried to analyze the physical reasons behind the light-harvesting complexes.
Our analysis is a retrospective antennas design, which automatically has the structural information enforced, as the antennas have already been made by nature.
The analysis conforms to the current understanding of light-harvesting antennas in that we can explain
\begin{itemize}
\item The functional roles of the notch at the light harvester,
\item The functional role of the PufX/W,
\item The functional role of the special pair, 
\item Reasons that the cross section of the light harvester must not be circular, 
\item A reason that the light harvester must not be spherical,
\item Reasons for the use of dielectrics instead of conductors to make the light harvester,
\item A mechanism to prevent damage from excess sunlight,
\item An advantage of the dimeric form, and
\item Reasons for the modular design of nature.
\end{itemize}

Based upon our analysis a mechanism for dimerization is provided.
We propose that more dimeric than monomeric complexes will appear in cells cultured under intense sunlight.



\bibliography{library} 

\end{document}